\documentstyle[12pt]{article}

\begin{document}
{\sf \begin{center} \noindent {\Large \bf Geometry of Liquid Crystals and the alignment of HIV-1 virus helical filaments }\\[3mm]

by \\[0.3cm]

{\sl L.C. Garcia de Andrade}\\

\vspace{0.5cm} Departamento de F\'{\i}sica
Te\'orica -- IF -- Universidade do Estado do Rio de Janeiro-UERJ\\[-3mm]
Rua S\~ao Francisco Xavier, 524\\[-3mm]
Cep 20550-003, Maracan\~a, Rio de Janeiro, RJ, Brasil\\[-3mm]
Electronic mail address: garcia@dft.if.uerj.br\\[-3mm]
\vspace{2cm}
\end{center}
\paragraph*{}
 Helical geometry of liquid crystals (LC) \cite{1} has been proved very useful in the study of material science through the investigation
 of the so-called topological defects , disclinations linked in general to curvature and dislocations connected with torsion of
 three-dimensional space. More recently R. Kamien \cite{2} has recalled the fact that curved crystals in materials behave geometrically
 as the curved spacetime of Einstein general relativity , which keeps bound the stars galaxies and the universe through gravitational
 force. Molecular attractive forces in biology and physics on the microscopic level would play the role of the gravitation. One
 of the main useful applications of these helical geometries is on the investigation of large macromolecules such as the double helix
 structure of DNA \cite{3}. The main ingredient in the investigation of curved liquid crystals is the so-called torsion of the crystal
 structure, as shown recently by David Nelson and his Harvard group \cite{4}. In 1998, Normann Watts and his group \cite{5} have shown
 that LC can be useful in the experimental detection of the helical geometry of the HIV-1 Rev filaments. In their experiment
 the LC used keeps its integrity or stability as a fundamental step in the experiment. This allows us to investigate recently
 the Lagrangean or geometrical stability of curved and torsioned (or twisted) LC \cite{6} which by computed the
 geodesic deviation of particles around the liquid crystals, which could be the moving of HIV molecules, diverge when the
 sectional curvature is negative, which indicates instability while the stability is connected to the positiveness of curvature.
 This simple mathematical technique applied to LC allows us to stablished the integrity or not of the LC used to test HIV viruses
 structure. It is not difficulty to show from the geometrical viewpoint that the stability or integrity of LC depends upon the
 pitch of the helix of double twisted blue phases of the crystals.
 This type of scientific activity may bind together several
 specialists of different areas of science, such as solid state
 physicists, soft  matter condensed matter physicists, DNA and
 molecular biology and chemical engineers and biochemists. Another
 type of the application of these stability Lagrange investigations
 in on the plasma astrophysics where the solar filaments \cite{7} undergo the
 same helical geometry that seems to be unfolded now in the realm of
 previously hidden HIV-1 retrovirusus investigation.

\end{document}